\documentclass[12pt,preprint]{elsarticle}

\usepackage[T1]{fontenc}
\usepackage[utf8]{inputenc}

\usepackage{float}
\usepackage{textcomp}
\usepackage{amsmath}
\usepackage{amssymb}
\usepackage{graphicx}

\makeatletter
\def\ps@pprintTitle{%
  \let\@oddhead\@empty
  \let\@evenhead\@empty
  \let\@oddfoot\@empty
  \let\@evenfoot\@empty
}

\begin{document}

\title{Stylized Facts and Their Microscopic Origins: Clustering, Persistence,
and Stability in a 2D Ising Framework}

\author{Hernán E. Benítez$^{a}$, Claudio O. Dorso$^{a,b}$}

\address{$^{a}$Departamento de Física, Facultad de Ciencias Exactas y Naturales,
Universidad de Buenos Aires, Pabellón I, Ciudad Universitaria, 1428
Buenos Aires, Argentina\\
 $^{b}$Instituto de Física de Buenos Aires (IFIBA), Universidad de
Buenos Aires, Pabellón I, Ciudad Universitaria, 1428 Buenos Aires,
Argentina}
\begin{abstract}
The analysis of financial markets using models inspired by statistical
physics offers a fruitful approach to understand collective and extreme
phenomena\citep{Bornholdt_2003,Mantegna,Sornette} In this paper,
we present a study based on a 2D Ising network model where each spin
represents an agent that interacts only with its immediate neighbors
plus a term reated to the mean field \citep{Binder,Bornholdt_2001}.
From this simple formulation, we analyze the formation of spin clusters,
their temporal persistence, and the morphological evolution of the
system as a function of temperature\citep{Castellano,Stauffer_2004}.
Furthermore, we introduce the study of the quantity$1/2P\sum_{i}|S_{i}(t)+S_{i}(t+\Delta t)|$,
which measures the absolute overlap between consecutive configurations
and quantifies the degree of instantaneous correlation between system
states. The results show that both the morphology and persistence
of the clusters and the dynamics of the absolute sum can explain universal
statistical properties observed in financial markets, known as stylized
facts\citep{Bornholdt_2001,Lux_1999,Stanley_2008}: sharp peaks in
returns, distributions with heavy tails, and zero autocorrelation.
The critical structure of clusters and their reorganization over time
thus provide a microscopic mechanism that gives rise to the intermittency
and clustered volatility observed in prices\citep{Bornholdt_2001,Sornette}. 
\end{abstract}
\maketitle

\section{Introduction}

The study of complex systems has found in statistical physics a conceptual
framework capable of offering profound analogies with economic and
financial phenomena\citep{Cont_Bouchaud_2001,Mantegna,Sornette}.
The Ising model allows us to describe phase transitions in magnetic
materials\citep{Ising,Stanley_197}, when we associate spins with
agents. When spins are associated to trading agents, this model has
become a useful paradigm for exploring collective behavior in markets\citep{Bornholdt_2001,Lux_1999,Stauffer_2007}.
The interaction between spins, their organization into clusters (sets
of spins that are immediate neighbors and have the same orientation),
and the influence of \textquotedbl temperature\textquotedbl{} allow
us to establish correspondences with the way in which financial agents
coordinate, generate trends, and navigate episodes of instability\citep{Bornholdt_2001,Castellano,Cont_2001}.

~

We analyze how cluster sizes and their dependence on temperature reflect
different relations between agents: extreme coordination during periods
of consensus, the emergence of heavy tails in critical situations,
and relative independence in stable scenarios\citep{Bornholdt_2003,Glauber,Stanley_2008}.
The formation and persistence of spin clusters constitute the starting
point for interpreting the mechanisms of collective order and disorder
that underlie price behavior\citep{Stauffer_2007,Yakovenko}.

~

Along with the morphological study, we introduce a new dynamic magnitude,
the microscopic stability factor (MSF) defined as:

\begin{equation}
MSF\text{=}\frac{1}{2P}\sum_{i}\left|S_{i}(t)+S_{i}(t+\Delta t)\right|
\end{equation}

which quantifies the absolute equivalence between consecutive configurations
of the system. This parameter measures the degree of instantaneous
correlation between the states of the spin set and allows us to characterize
the structural stability of the system over time, where the index
i denotes the spin. Each term $\mid Si(t)+Si(t+\Delta t)\mid/2$ equals
1 if the spin did not change between times $t$ and $t+\Delta t$,
and 0 if it did, where P is the number of sites in the network. Therefore,
the MST measures the fraction of spins that remain in the same state
between two consecutive configurations. High values of this quantity
reflect coherence and persistence between configurations---analogous
to trend or consensus phases in markets---while low values indicate
rapid and chaotic reconfigurations, associated with periods of high
volatility.

~

The morphology of the clusters, their temporal persistence and the
behavior of the MST clusters provide a microscopic framework that
allows us to explain the main statistical properties of financial
markets, known as stylized facts: abrupt peaks in returns, distributions
with heavy tails and zero autocorrelation.
~
The system is built on a square network of size $NxN$ with periodic
boundary conditions, and a total of $P=N^{2}$ sites. At each site,
a spin is located that can only take the values +1 or \textminus 1,
associated with the agent's decisions: +1 corresponds to a purchase
order, while \textminus 1 represents a sale order. Each agent updates
its state probabilistically according to a local field, applying Glauber
dynamics. The local field $h_{i}(t)$ describes the influence on spin
\emph{i} of the rest of the network set.

~

The dynamics are defined by the following equation comprising two
terms, one related to the interaccion with de 4 nearest neighbours
and the other with the mean field associated to de rest of the spins:

\begin{equation}
h_{i}(t)=\sum_{\langle i,j\rangle}J\,s_{j}(t)-\alpha\,s_{i}(t)\,\left|M(t)\right|
\end{equation}

with

\begin{equation}
M(t)=\frac{1}{P}\sum_{i=1}^{P}s_{i}(t)
\end{equation}

The temporal evolution of the system is stochastic and is implemented
using Glauber dynamics with random, serial, and asynchronous spin
updates\citep{Glauber}. This realistically represents the heat bath,
whereby each spin $S_{i}(t)$ is updated stochastically according
to the value of the local field affecting it. The probability that
a spin takes the value +1 in the next step is:

\begin{equation}
P\big(s_{i}(t+1)=+1\big)=\frac{1}{1+\exp\!\left[-2\beta\,h_{i}(t)\right]}
\end{equation}

and, in a complementary manner,

\begin{equation}
P\big(s_{i}(t+1)=-1\big)=1-P\big(s_{i}(t+1)=+1\big)
\end{equation}

Here, $\beta$ is a parameter (which in statistical mechanics is associated
with the temperature of the system, $\beta=1/T$) that regulates the
transition probabilities of the spins, between the states +1 and -1.
In summary, Glauber dynamics translates the influence of the local
field into a probability of state change, allowing the spins to evolve
over time depending on their environment.

~

Considering the terms of equation (2):

1. Local interaction: the first term describes the tendency of each
spin to align with its nearest neighbors $\left\langle i,j\right\rangle $.
This contribution is ferromagnetic and reproduces the behavior of
the traditional Ising model $J>0$. In all experiments, $J=1$ is
taken.

2. Minority behavior: the second term introduces a dynamic of opposite
sign, which favors the change of state and is interpreted as a “contrary
strategy” effect\citep{Bornholdt_2001,Takaishi}. Controlled by the
parameter $\alpha$ (is constant over time), this mechanism introduces
fluctuations that break the overall consensus.In all experiments,
the same value $\alpha=4$ is used.

~

\section{Spin clusters as market coalitions}

The dynamics of the Ising model lead to the formation of aligned spin
clusters. Drawing inspiration from fragment recognition methods in
molecular dynamics\citep{Chernomoretz,Strachan_Dorso}, a cluster
is formally defined as a set of spins \ensuremath{i},\ensuremath{j},\ensuremath{k},…
that simultaneously satisfy a connectivity and correlation condition.

~

In the simplest approximation, clusters are detected by considering
only the first neighbors of each site in the square lattice. Thus,
two spins located at positions $\vec{r_{i}}$ and $\vec{r_{j}}$ belong
to the same cluster \ensuremath{C} if:

\begin{equation}
i \in C \iff \exists\, j \in C \;:\; |r_i - r_j| = a \;\wedge\; \sigma_i = \sigma_j ,
\end{equation}

where \ensuremath{a}=1, and is the distance between first neighbors
in the lattice (the mesh spacing), and $\sigma_{i}$,$\sigma_{j}\in\left\{ +1,-1\right\} $
are the spin values. In other words, two spins are part of the same
cluster if they are aligned and connected by a continuous chain of
first neighbors with the same orientation.

~

These spin clusters can be interpreted as coalitions of traders taking
the same position (buyers or sellers). The size and distribution of
the clusters provide an indicator of the degree of coordination in
the market.

~

• Large clusters represent massive consensus (low volatility).

• Small and distributed clusters reflect the coexistence of multiple
opinion groups (high volatility).

\subsection{Morphology of the system as a function of $1/\beta$ and for $\alpha=4$.}

~

Network dynamics allow us to characterize different morphological
regimes $1/\beta$ to with $\alpha=4$ (see appendix for $\alpha=0$).
Alpha acts as a control between collective order and disorder. In
the analogy with markets, $1/\beta$ regulates the balance between
social imitation (tendency to form large clusters) and heterogeneity
in individual decisions (small clusters).

\subsubsection{$1/\beta$ $\ll1/\beta_{c}$: two phases.}

~

In this regime the system exhibits a strong tendency to form two dominant
clusters, a product of the second term in equation 2. Here, $1/\beta_{c}$
represents the critical temperature of the Ising system, that is,
the point where the phase transition occurs between the disordered
(paramagnetic) and ordered (ferromagnetic) states. Spin correlations
become long-range, and spontaneous magnetization (equation 3) appears
or disappears. The non-zero magnetization reflects that most spins
remain aligned, generating large and persistent domains. Morphologically,
the lattice is organized into a few large clusters that remain stable
over time (see Figure 1).

\begin{figure}[H]
\begin{raggedright}
\includegraphics[width=\linewidth]{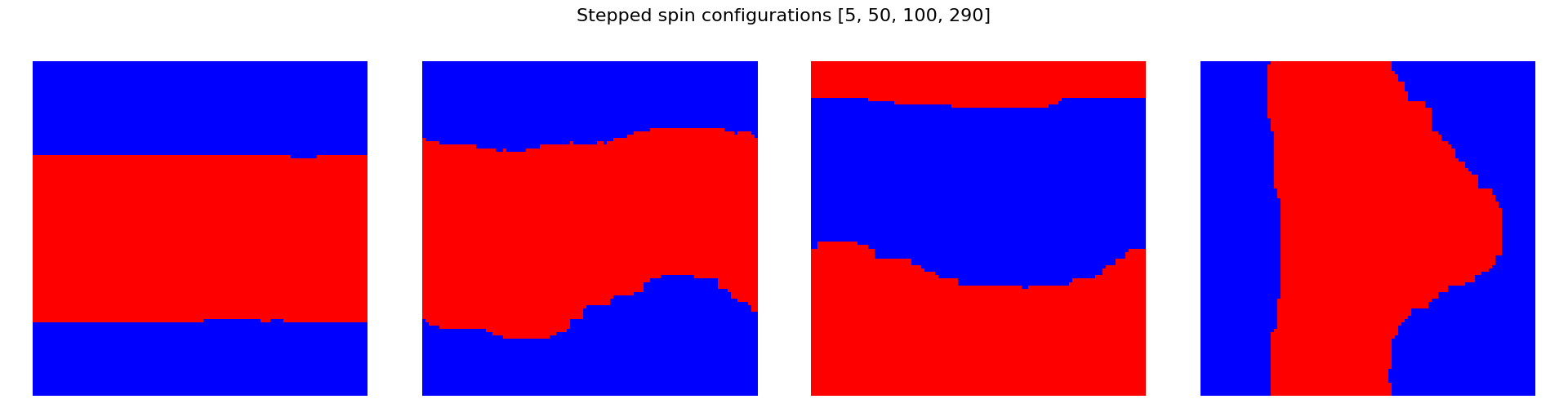} 
\par\end{raggedright}
\caption{Spin configurations obtained from the dynamics of the two-dimensional
Ising model at $1/\beta=0.5$ and $\alpha=4$. Each configuration
corresponds to a sample taken every 1000 time steps, with the instants
5k, 50k, 100k, and 290k being arbitrarily selected. At this $1/\beta$,
the system exhibits a coexistence of large and small domains that
evolve slowly over time. Although the general structure of the clusters
is preserved, fluctuations and gradual shifts of the domain boundaries
are observed, reflecting a slow relaxation dynamic toward more ordered
states (Color online).}
\end{figure}

~

At this value of $1/\beta$ , the system exhibits a slow relaxation
dynamic: although the overall cluster structure is preserved, fluctuations
and gradual shifts in the domain boundaries are observed. This is
reflected in the cluster size distribution (Figure 2), where large
and small clusters coexist, indicating that the consensus among the
agents (spins) is not absolute.

~

In financial terms, these scenarios can be interpreted as periods
in which trends predominate, but with room for divergences: large
groups of agents adopt similar positions, generating directed movements,
while smaller clusters represent local dissent that can slowly alter
the dynamics. The coexistence of large and small domains thus reflects
partial stability with moderate persistence of market structures,
unlike extreme situations where a single cluster dominates the system.
In this context, clusters are delimited by interfaces that possess
an associated energy, called interface energy. For a spin to be reversed,
it must overcome the energy barrier imposed by this interface\citep{Hongler,Southern}.
At low temperatures, interface energy is considerable compared to
thermal energy, making spin changes difficult: reversing a spin involves
breaking part of the interface, thus requiring a significant amount
of energy (see Figure 2).

\begin{figure}[H]
\begin{centering}
\includegraphics[width=\linewidth]{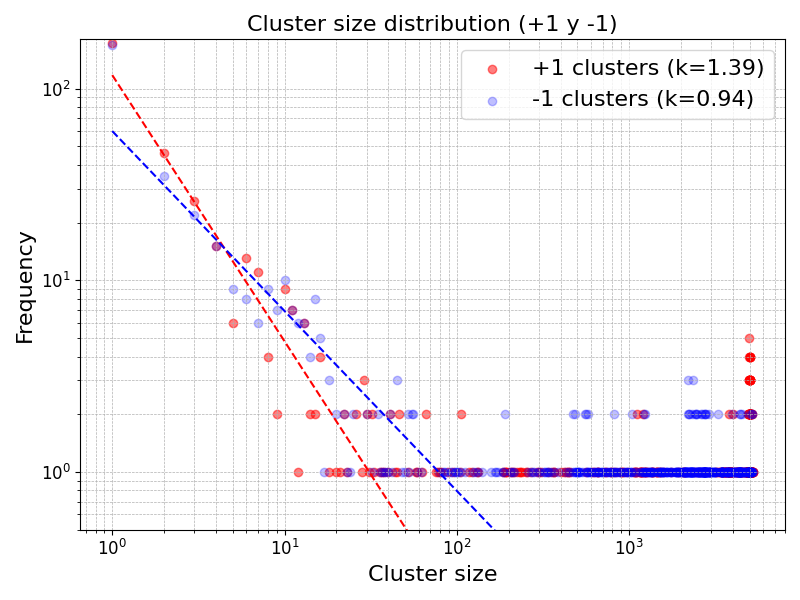} 
\par\end{centering}
\caption{Below $1/\beta$, the +1 clusters (red) exhibit a steeper slope in
the distribution, indicating a predominance of small clusters, while
the -1 clusters (blue) show a heavier distribution, with a higher
probability of finding large clusters. The dispersion of the points
suggests oscillations in cluster formation, reflecting the active
dynamics of the domains at this temperature (Color online).}
\end{figure}

\subsubsection{$1/\beta_{c}$: instability and clusters of all scales}

~

Near the “critical temperature” ($1/\beta=T\approx Tc=2.26$), the
system's morphology changes dramatically. Cluster sizes no longer
have a characteristic scale, and their distribution follows power
laws with exponents on the order of $\approx1.9$. This reflects the
coexistence of domains of all magnitudes, from small local aggregates
to clusters that encompass a significant fraction of the network (see
Figure 3). Persistence decreases markedly: large clusters form and
dissolve rapidly, reflecting highly fluctuating dynamics. In the context
of markets, this critical phase is interpreted as a state of high
instability: extreme returns become more probable, and systemic susceptibility
reaches its peak.

\begin{figure}
\includegraphics[width=\linewidth]{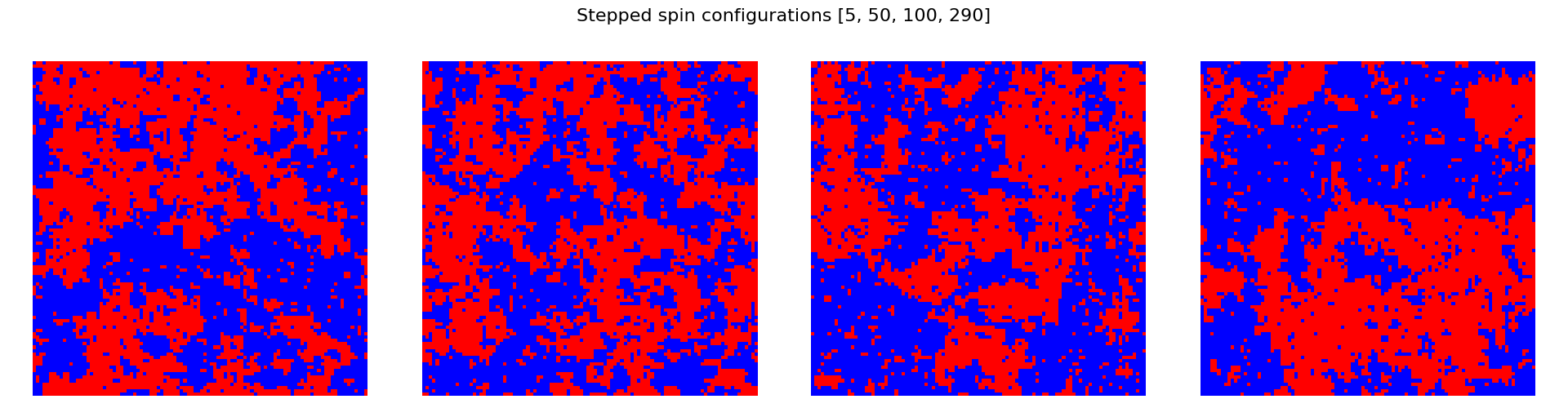}

\caption{The cluster size distribution at $1/\beta=2.2$ exhibits a power-law
behavior, indicating the absence of a characteristic scale. In this
regime, the dynamics are highly variable and allow for the simultaneous
coexistence of clusters of all sizes, from the smallest to the largest
(Color online).}
\end{figure}

The correlation extends throughout the entire system, so a local disturbance
can amplify to trigger a global reordering. In the market, this translates
into episodes of crisis or high volatility, where extreme returns
are more likely and financial systems become particularly vulnerable
to systemic risk. The loss of typical scales in cluster dynamics is
reflected in the loss of characteristic scales in returns, a phenomenon
that in economics manifests itself through distributions with heavy
tails\citep{Cont_2001,Cont_Bouchaud_2001}. See Figure 4.

\begin{figure}[H]
\begin{centering}
\includegraphics[width=\linewidth]{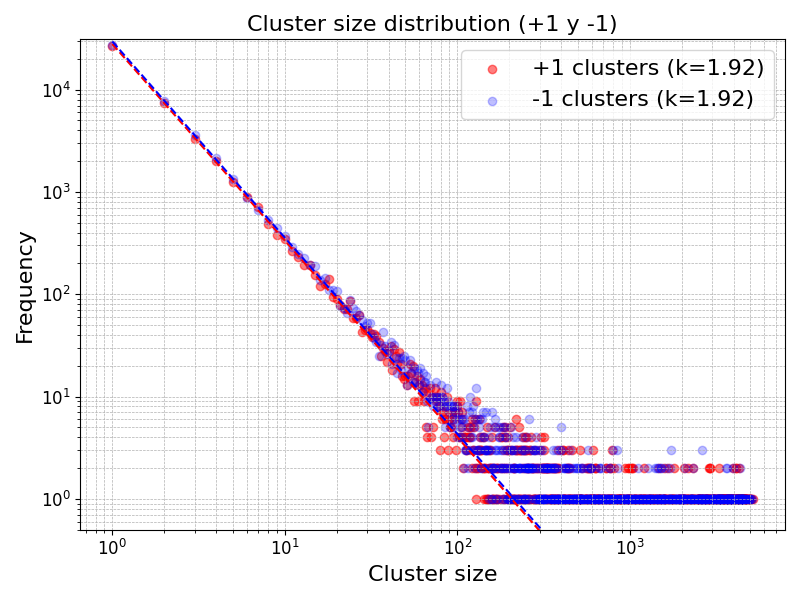} 
\par\end{centering}
\caption{Close to the critical \textquotedblleft temperature\textquotedblright{}
$1/\beta=T=2.2$ ($1/\beta_{c}=T_{c}=2.26$). A critical state is
observed in the Ising model (or any system with competing domains).
None of the phases dominates. The cluster size distribution follows
a power law with an exponent of \textasciitilde 1.9, typical of criticality
(Color online).}
\end{figure}

\subsubsection{$1/\beta\gg1/\beta_{c}$: disorder and independence}

~

When the value of $1/\beta$ is high ($1/\beta\gg1/\beta_{c}$), the
system's morphology is characterized by fragmentation into small,
ephemeral clusters. Thermal agitation dominates over ferromagnetic
interaction, and the spins fluctuate almost independently (see Figure
5). In this scenario, cluster size distributions approximate Gaussian
shapes, and persistence is reduced to practically zero. The analogy
with markets is a stable yet fluid environment, in which agents' decisions
are dispersed and no significant large-scale correlations are generated.

\begin{figure}[H]
\begin{raggedright}
\includegraphics[width=\linewidth]{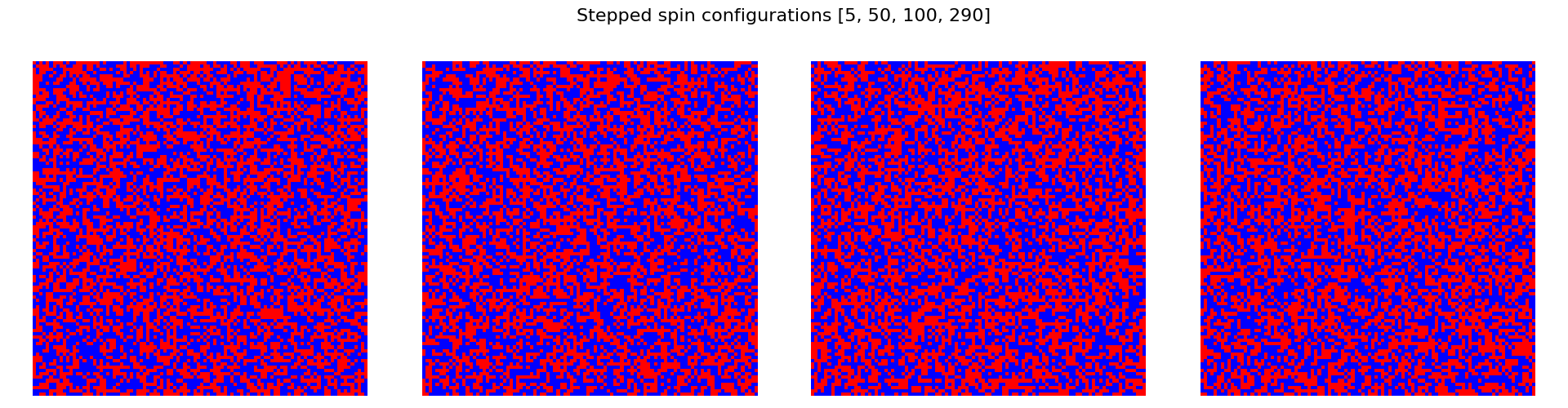} 
\par\end{raggedright}
\caption{At high $1/\beta=10$, the clusters remain small and unstable. Intense
thermal agitation causes them to constantly form and dissolve, generating
configurations dominated by small, ephemeral clusters, as shown in
the figure (Color online).}
\end{figure}

~

When the temperature is high, the spins fluctuate almost independently,
and the clusters are small, distributed with profiles close to normality.
In this sense, the analogy in the markets is a state of relative calm,
in which agents act mostly autonomously and the system does not exhibit
significant correlations; that is, they do not act in a coordinated
manner. In these scenarios, the market resembles a liquid and efficient
environment, where prices reflect dispersed information and returns
approach Gaussian behavior. See Figure 6.

\begin{figure}[H]
\begin{centering}
\includegraphics[width=\linewidth]{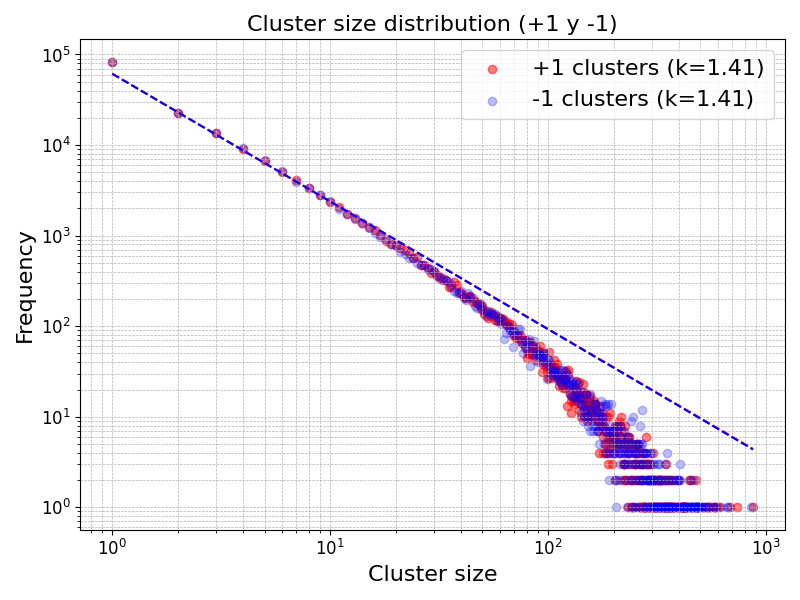} 
\par\end{centering}
\caption{High $1/\beta=10$. The largest clusters are 6\% of the network size
(100x100). The distribution has clearly shifted towards relatively
small clusters. Large clusters do not coexist at these temperatures,
indicating that the consensus is dispersed in small clusters relative
to the total network size (Color online).}
\end{figure}

\section{Cluster persistence at consecutive times (t and t+1).}

It is important to know whether the agents that make up a cluster
at a certain time remain aligned at a later time, that is, still forming
part of the same cluster. This is called cluster persistence. We then
analyze the persistence of the largest clusters of spins +1 and \textminus 1
at different temperatures ($1/\beta=0.5$, $1/\beta=2.2$ and $1/\beta=10$).
A clearly temperature-dependent behavior is observed. Following the
definition introduced in the context of nuclear fragmentation \citep{Chernomoretz},
persistence is measured through short-time persistence (STP). For
a cluster $C_{it}$of size $N_{it}$ at time $t$, the largest subset
of particles $N_{max}^{t+\triangle t}$ that remain bound is identified
at time $t+\Delta t$ This allows us to define:

\begin{equation}
{STP}_{d}=\frac{N_{\max_{i}}^{t+\Delta t}}{N_{i}^{t}},\qquad{STP}_{i}=\frac{N_{\max_{i}}^{t+\Delta t}}{N_{i}^{t+\Delta t}}
\end{equation}

Global persistence results from the average:

\begin{equation}
STP(t,\Delta t)=\left\langle \left\langle \frac{STP_{d}(t,\Delta t)+STP_{i}(t,\Delta t)}{2}\right\rangle _{j}\right\rangle _{e}
\end{equation}

where the first average is calculated over all clusters weighted by
their size and the second over an ensemble of configurations. Thus,
a value of $STP\approx1$ indicates that the clusters are stable between
time steps, while $STP\ll1$ reflects significant disintegration or
reconfiguration.

~

At low $1/\beta$, short-term persistence remains very close to 1.
This means that the larger clusters of +1 and \textminus 1 spins almost
entirely retain their microscopic composition between successive steps,
with only minor local in time fluctuations. Low $1/\beta$, in relation
to spin interaction, stabilizes the configurations and prevents cluster
fragmentation. The dynamics are slow and dominated by constant domains
that remain virtually unchanged over time (see Figure 7).

\begin{figure}[H]
\begin{centering}
\includegraphics[width=\linewidth]{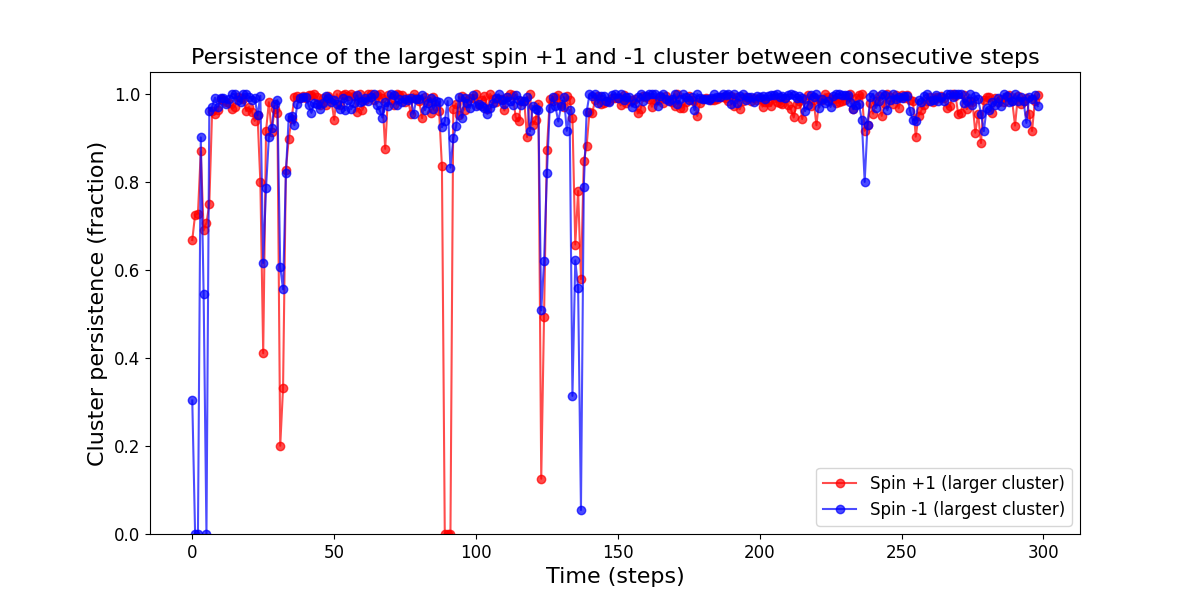} 
\par\end{centering}
\caption{Persistence of the largest cluster of spins +1 and \textminus 1 in
the Ising model for regimes of $1/\beta=0.5$ (stable order) (Color
online).}
\end{figure}

In the intermediate regime, near $1/\beta_{c}$, the persistence of
the largest clusters drops markedly, reaching typical values in the
range $0\lesssim STP\lesssim0.6\text{–}0.7$. This reflects the
rapid and frequent changes in the dominant cluster composition: clusters
are constantly forming and dissolving. Here, thermal energy competes
with spin interactions(i.e. spins are flipping at a rather high rate),
generating instability in the configurations. As a result, the dynamics
become highly fluctuating and chaotic, with large-scale loss of order
(see Figure 8).

\begin{figure}[H]
\begin{centering}
\includegraphics[width=\linewidth]{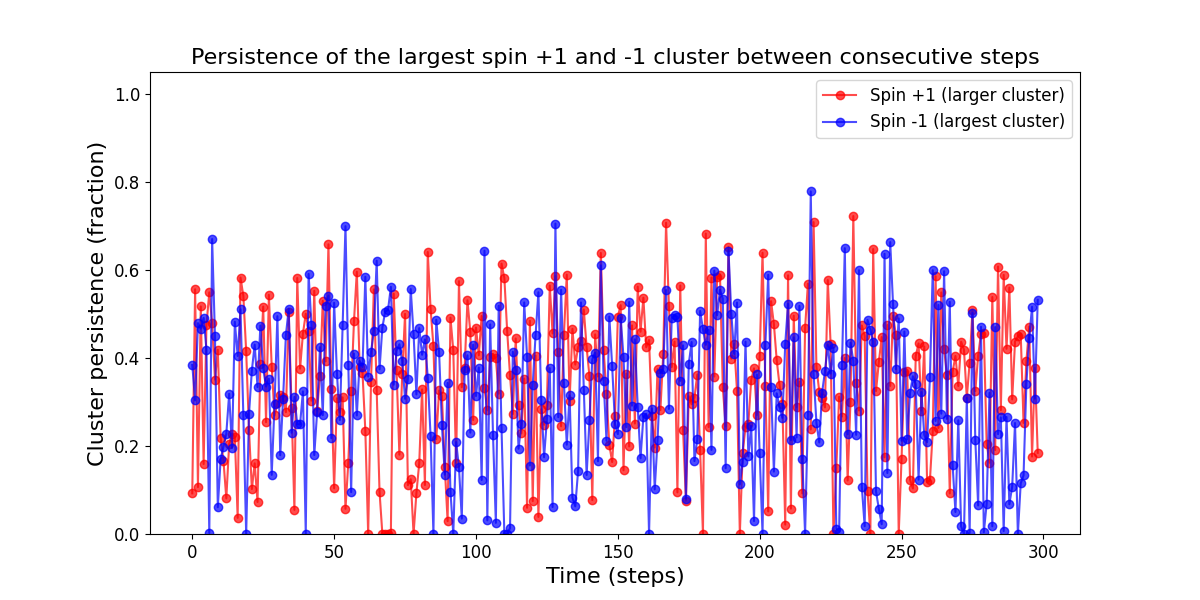} 
\par\end{centering}
\caption{Persistence oscillates around $1/\beta=2.2$ (close to $1/\beta_{c}$),
and is observed reflecting a transition from an ordered state to an
unordered one (Color online).}
\end{figure}

At high $1/\beta$, persistence reaches small STP values. Even in
the rare cases where relatively large clusters form, they disintegrate
almost immediately. Thermal agitation completely dominates and renders
spin interactions irrelevant. Consequently, stable clusters are not
sustained, and the dynamics are governed by total disorder when agents
switch from one opinion to the contrary (see Figure 9).

\begin{figure}[H]
\begin{centering}
\includegraphics[width=\linewidth]{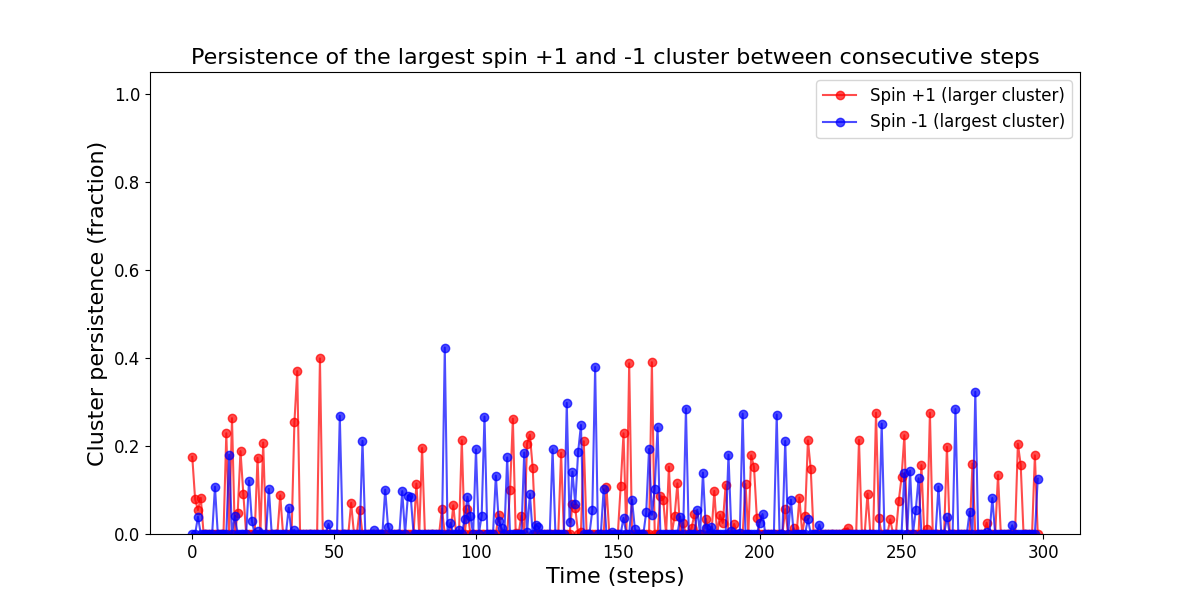} 
\par\end{centering}
\caption{For $1/\beta=10$ (complete disorder). It is observed how persistence
decreases with increasing $1/\beta$, reflecting the transition from
an ordered state to an almost completely disordered one. Clusters
don't even have time to form before they disintegrate (Color online).}
\end{figure}

\section{Microscopic Stability Factor (MSF)}

Equation 1 defines the microscopic stability factor (MSF), which measures
the temporal persistence of the complete spin configuration between
two consecutive states. The MSF ranges from 0 (complete reconfiguration,
minimum MSF) to 1 (fully persistent configuration). The evolution
of the MSF was studied by varying $1/\beta$, from an initial value
of 10 to 0.1 ($0.1\leq1/\beta\leq10)$. The simulation consisted of
$2.10^{6}$ time steps, and the first $25.10^{3}$ steps were used
to \textquotedbl thermalize\textquotedbl{} the lattice. This procedure
allowed observation of how the microscopic stability factor develops
progressively as the system cools, capturing the transition between
disordered and ordered regimes. Two cases were studied: $\alpha=0$
and $\alpha=4$.

\subsection{Case $\alpha=0$: absence of global coupling (see equation 2).}

~

In this case, we return to the usual Ising model. Figure 10 shows
the evolution of the MSF. It increases smoothly as $1/\beta$ decreases,
showing the transition from a slightly disordered to an ordered regime.
At values of $4\leq1/\beta\leq10$, the MST remains close to 0.75,
indicating a high rate of spin reconfiguration and, therefore, weak
morphological memory: clusters are constantly fragmenting and restructuring.
As the system cools, fluctuations decrease and domains begin to consolidate,
increasing the fraction of spins that retain their state between consecutive
configurations. This evolution reflects a self-organizing process,
where stability emerges from local interactions without the need for
global feedback.

\begin{figure}[H]
\begin{centering}
\includegraphics[width=\linewidth]{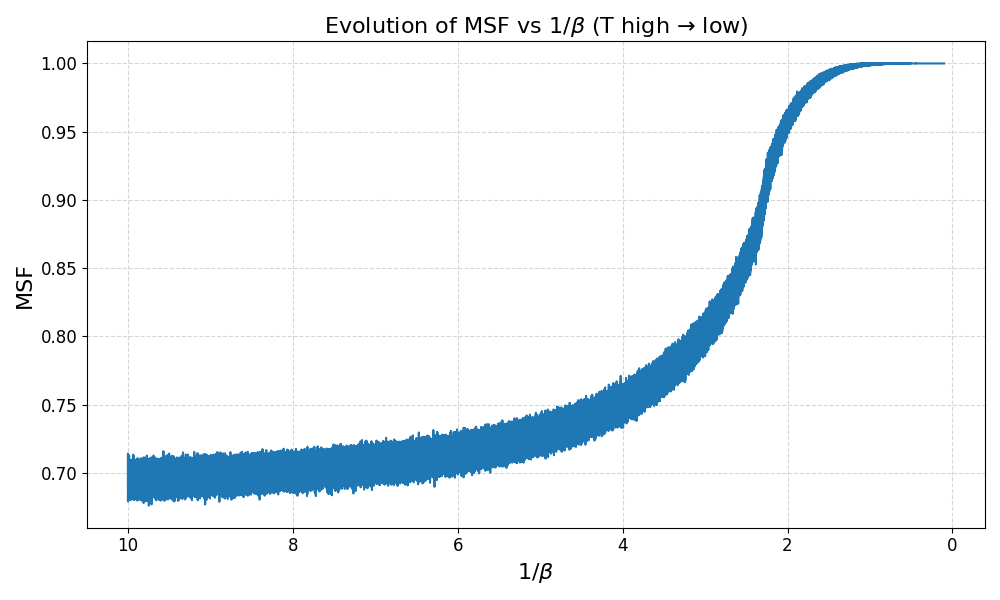} 
\par\end{centering}
\caption{Evolution $MSF=\frac{1}{2P}\sum_{i}\left|S_{i}(t)+S_{i}(t+\Delta t)\right|$
for $\alpha=0$ as a function of $1/\beta$. The system transitions
from a disordered to an ordered regime as $1/\beta$ decreases, demonstrating
a monotonic growth of structural persistence (Color online).}
\end{figure}

.

\subsection{Case $\alpha=4$: with global coupling (see equation 2).}

~

When $\alpha$ is different from 0, the energy term acquires an additional
global contribution that couples the magnetization of the entire system,
effectively modifying the stability landscape. This extra term introduces
a large-scale influence that competes with purely local interactions.
Figure 11 shows a different behavior. Global coupling introduces a
feedback term that penalizes strongly ordered states, generating a
competition between the local tendency toward order and the global
force of destabilization. Consequently, the MSF curve exhibits intermittent
oscillations at low $1/\beta$ values, signaling alternation between
phases of high and low morphological memory. The system enters a dynamic
regime where clusters form and dissolve abruptly, reflecting the coexistence
of stable domains and sudden collective reorganizations.

\begin{figure}[H]
\begin{centering}
\includegraphics[width=\linewidth]{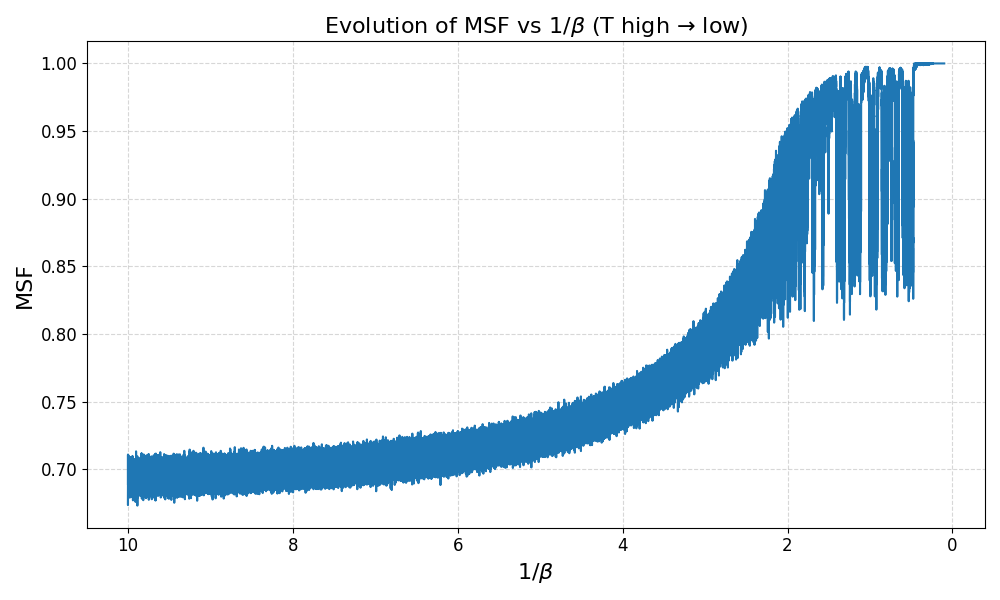} 
\par\end{centering}
\caption{MSF for $\alpha=4$. Global feedback generates an intermittent regime
with alternation between stable phases and sudden reorganizations
(Color online).}
\end{figure}

\subsection{Returns}

~

The return, $R(t)$is defined as:

\begin{equation}
R(t)=\ln[M(t)]-\ln[M(t-1)])
\end{equation}

wheare the price is associated with the magnetization $M(t))$ \citep{Bornholdt_2001,Cont_2001,Cont_Bouchaud_2001,Mantegna}.

~

Figure 12 shows the time evolution of the returns for the case $\alpha=0$.
A clear attenuation of the fluctuations is observed as soon as the
value approaches $1/\beta_{c}$. The system evolves, with the initially
large returns decreasing until they oscillate around values close
to zero. This strong transition indicates that the system reaches
a stationary regime dominated by a stability factor (MSF → 1). In
this phase, the clusters remain stable, and most spins maintain their
orientation between consecutive configurations. From a financial perspective,
this dynamic is equivalent to a stable, low-volatility market where
agents maintain a persistent consensus. The returns are small, distributed
almost in a Gaussian way and without the presence of abrupt shocks,
reflecting a phase of structural order in the network.

\begin{figure}[H]
\begin{centering}
\includegraphics[width=\linewidth]{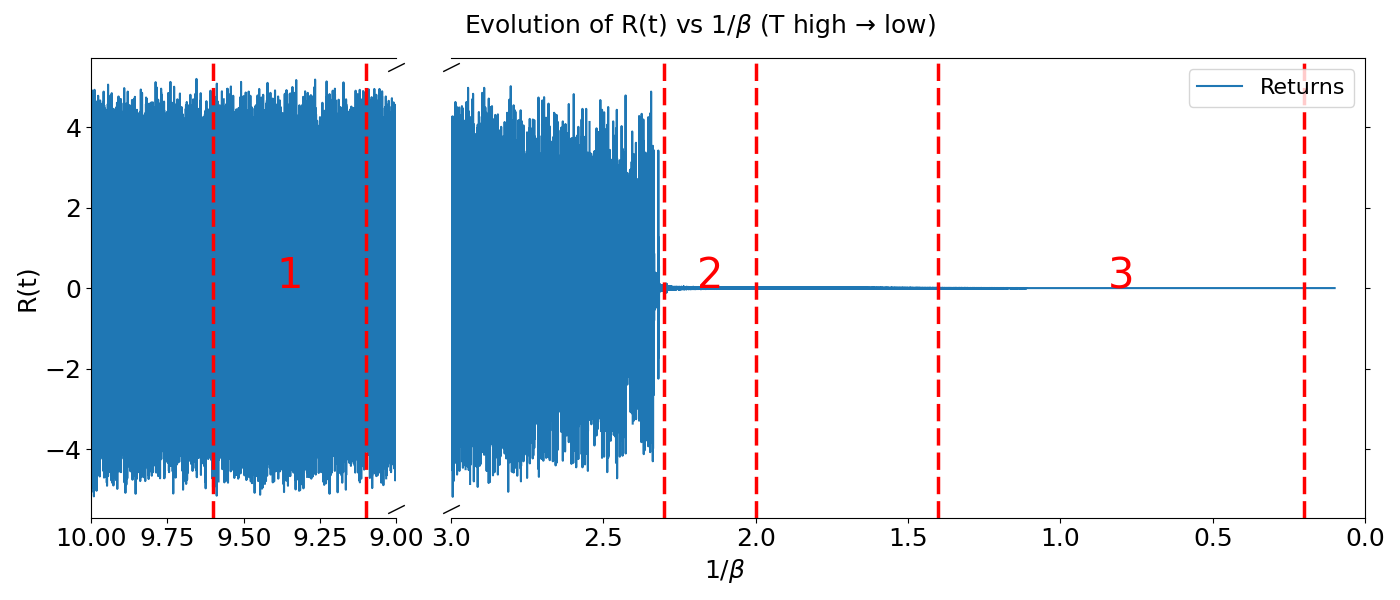} 
\par\end{centering}
\caption{Time evolution of returns for $\alpha=0$. The decrease in the amplitude
of the fluctuations indicates a regime of high morphological stability
and low volatility (Color online).}
\end{figure}

Figure 13 allows us to observe in greater detail the sectoral structure
of returns as a function of $1/\beta$

~ 
\begin{itemize}
\item In region 1 ($1/\beta>1/\beta_{c}$), the MSF is low ($\sim0.75$)
and clusters form and dissolve rapidly. This generates Gaussian returns,
without volatility clustering, analogous to a liquid market where
agents's decisions are independent. 
\item In region 2 ($1/\beta_{c}\approx2.26$), the MSF increases ($\sim0.9$)
and the returns are drastically attenuated, almost disappearing. The
system structure approaches a critical state in which morphological
persistence is high. 
\item Finally, in region 3 ($1/\beta<1/\beta_{c}$), the $MSF\sim1$ and
returns disappear. The clusters stabilize completely and the overall
magnetization remains constant, which in financial terms is equivalent
to a market in full consensus, without significant oscillations or
price fluctuations. 
\begin{figure}[H]
\includegraphics[width=\linewidth]{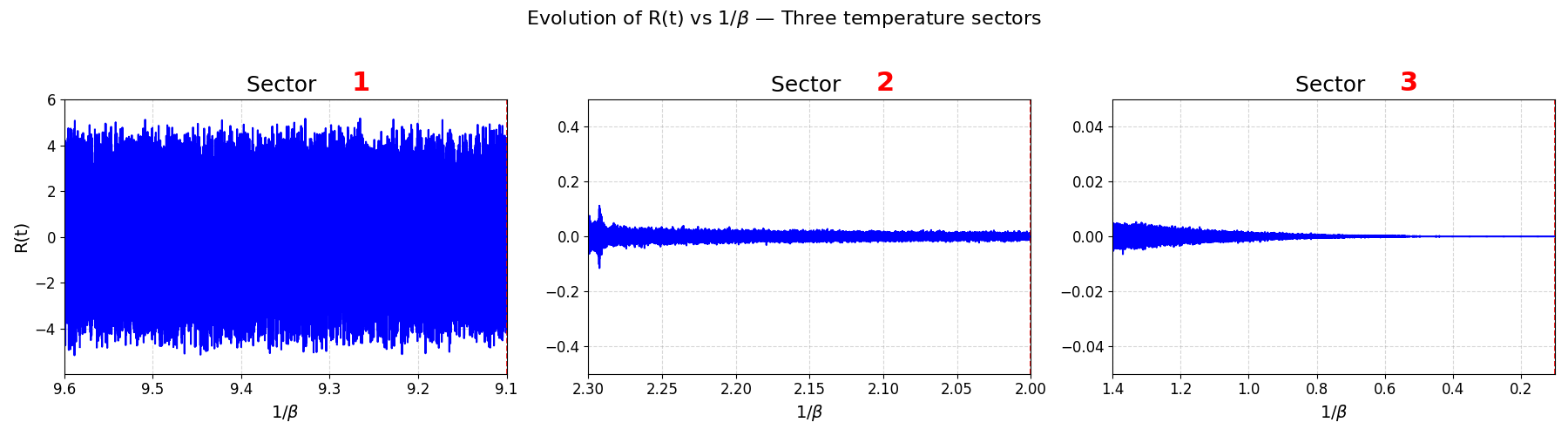}

\caption{Sectoral structure of returns for $\alpha=0$ as a function of $1/\beta$.
Each region reflects a distinct morphological regime: thermal disorder
(low memory), critical transition (intermediate memory), and collective
order (high memory), where returns attenuate until they disappear.
Please notice the change in scales (Color online).}
\end{figure}
\end{itemize}
In contrast, Figure 14 shows the evolution of returns when the global
coupling is $\alpha\text{=4}$ In this case, returns exhibit marked
intermittency, with large-amplitude oscillations even in states close
to global order. The system alternates between periods of high and
low morphological memory, reflecting abrupt collective reorganizations
in which large spin domains simultaneously change state. From a financial
perspective, this dynamic reproduces the stylized facts of real markets:
prolonged phases of calm (high MSF, low volatility) interspersed with
episodes of high endogenous volatility (low MSF), generated not by
external shocks but by the system's own internal self-organization.

\begin{figure}[H]
\begin{centering}
\includegraphics[width=\linewidth]{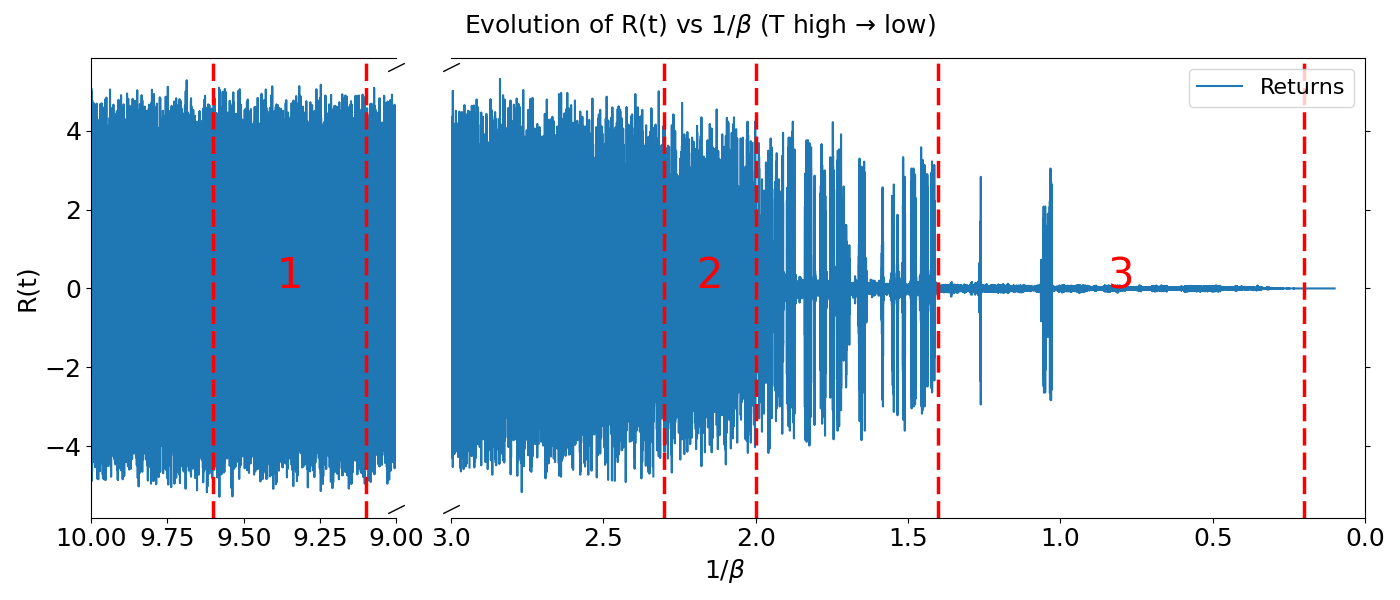} 
\par\end{centering}
\caption{Returns for $\alpha\text{ = 4}$ The presence of intermittent oscillations
and large amplitude peaks shows evidence of collective morphological
reorganizations, reflecting critical dynamics analogous to episodes
of extreme volatility in the markets (Color online).}
\end{figure}

Figure 15 presents a magnified view of these regimes. In high $1/\beta$
sectors, the MSF exhibits rapid oscillations that herald morphological
disorder and Gaussian returns. As the system cools, sudden drops in
MSF precede volatility spikes in returns, evidencing a loss of structural
coherence between spins and clustering of returns. In states of deep
order (high and stable MSF), the oscillations disappear, and returns
stabilize near zero. In financial terms, the MSF functions as a risk
indicator: its abrupt decline anticipates critical reorganizations
equivalent to financial crises or bubbles.

\begin{figure}[H]
\begin{raggedright}
\includegraphics[width=\linewidth]{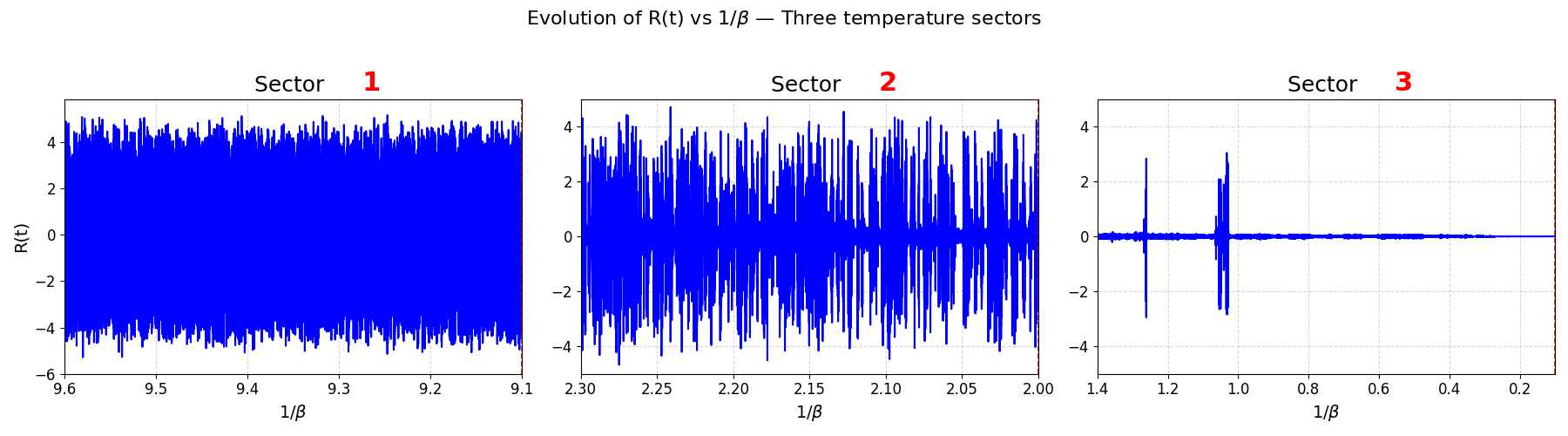} 
\par\end{raggedright}
\caption{Sectoral evolution of returns and MSF for $\alpha=4$ Intermittent
oscillations of morphological memory precede episodes of high volatility,
revealing the connection between structural disorder and extreme fluctuations
in returns (Color online).}
\end{figure}

\section{Relationship between clustering, persistence, and the microscopic
stability factor with the price series}

The global magnetization of the network, defined in (3), can be interpreted
as the analogue of an aggregate price of an asset. In this framework,
relative changes in magnetization represent financial returns see
Equation 9, establishing a bridge between the microscopic morphology
of the system and the temporal dynamics of prices \citep{Cont_2001}.
This approach allows for a direct link between the internal structure
of spin clusters and the macroscopic behaviors characteristic of financial
markets. In particular, it emphasizes that these clusters correspond
to groups of agents sharing the same orientation: the larger and more
compact a cluster is, the harder it becomes to reverse its state.

~

We begin by analyzing the two-dimensional Ising model, where the morphology
of spin clusters and their temporal persistence---measured by the
microscopic stability factor (MSF)---form the basis for understanding
the dynamics of magnetization (price). At low temperatures, spins
tend to align, forming large, stable clusters, which translates into
MSF values close to 1, indicating high structural coherence. In this
phase, overall magnetization varies slowly, and returns are small,
reflecting a stable market dominated by market consensus.

~

In the critical region (around $1/\beta_{c}$ \ensuremath{\approx}
2.26), the system reaches an unstable equilibrium between order and
disorder. Here, the MST fluctuates between intermediate values ($\sim0.8\text{–}0.9$),
indicating that the system partially preserves the spin configuration
between successive steps, but with frequent reorganizations. In this
regime, the clusters exhibit a hierarchical distribution that follows
power laws, implying the coexistence of multi-scale domains without
a characteristic length. These critical reorganizations manifest in
the returns as high-amplitude spikes and heavy tails, reflecting episodes
of high endogenous volatility generated by the system's own dynamics.

~

Conversely, at high temperatures, spin interactions weaken --- effectively
being overwhelmed by thermal fluctuations ---, clusters fragment,
and the MSF drops to values close to 0.7, indicating morphological
disorder and a rapid loss of persistence between configurations. In
this regime, returns exhibit a nearly Gaussian distribution, analogous
to a liquid market where agents act independently and fluctuations
are essentially random.

\section{Conclusions}

As we have show in the previous seccitions the introduction of the
MSF as defined in the Equation 1, allows as to better understand the
raison why the caracteristics behavior stylezed facts emerge. This
parameter measures the structural coherence of the system and allows
for the reinterpretation of stylized facts based on the microscopic
stability of the clusters. High MSF values indicate a stable morphology
with low volatility, while abrupt decreases in MSF reflect sudden
cluster reorganizations, equivalent to crises or endogenous bubbles
in the markets.

~

Taken together, these results establish a direct link between cluster
morphology, their temporal persistence, and stylized facts of financial
markets. In particular:

• Power-law return distributions arise from the intermittent reorganization
of clusters in the critical regime, where domains of all scales coexist
and the MSF exhibits intermediate fluctuations.

• Clustered volatility emerges from the oscillations of the MSF, which
alternates between phases of structural coherence (high memory) and
phases of breakdown (low memory).

• The absence of autocorrelation in returns is associated with the
rapid loss of morphological persistence between consecutive configurations,
analogous to the transitions between calm and crisis regimes in real
markets.

~

In summary, cluster morphology, temporal persistence, and the microscopic
stabilized factor (MSF) constitute the microscopic substrate that
explains macroscopic price phenomena. Large, persistent clusters generate
prolonged trends; critical configurations with intermediate memory
give rise to heavy tails and clustered volatility; and thermal fragmentation,
along with a low microscopic stability factor, corresponds to liquid
and Gaussian markets. Overall, this analysis confirms that the stability
and vulnerability of financial markets depend not only on price trajectories
but also on the underlying morphological organization and the system's
memory. In this sense, the Ising model offers a robust conceptual
framework for understanding how collective dynamics that give rise
to stylized facts in complex financial systems emerge, propagate,
and amplify. Figure 16 summarizes the relationship between clusters,
persistence, and the MSF to explain stylized price facts in financial
markets.

\begin{figure}[H]
\begin{centering}
\includegraphics[width=\linewidth]{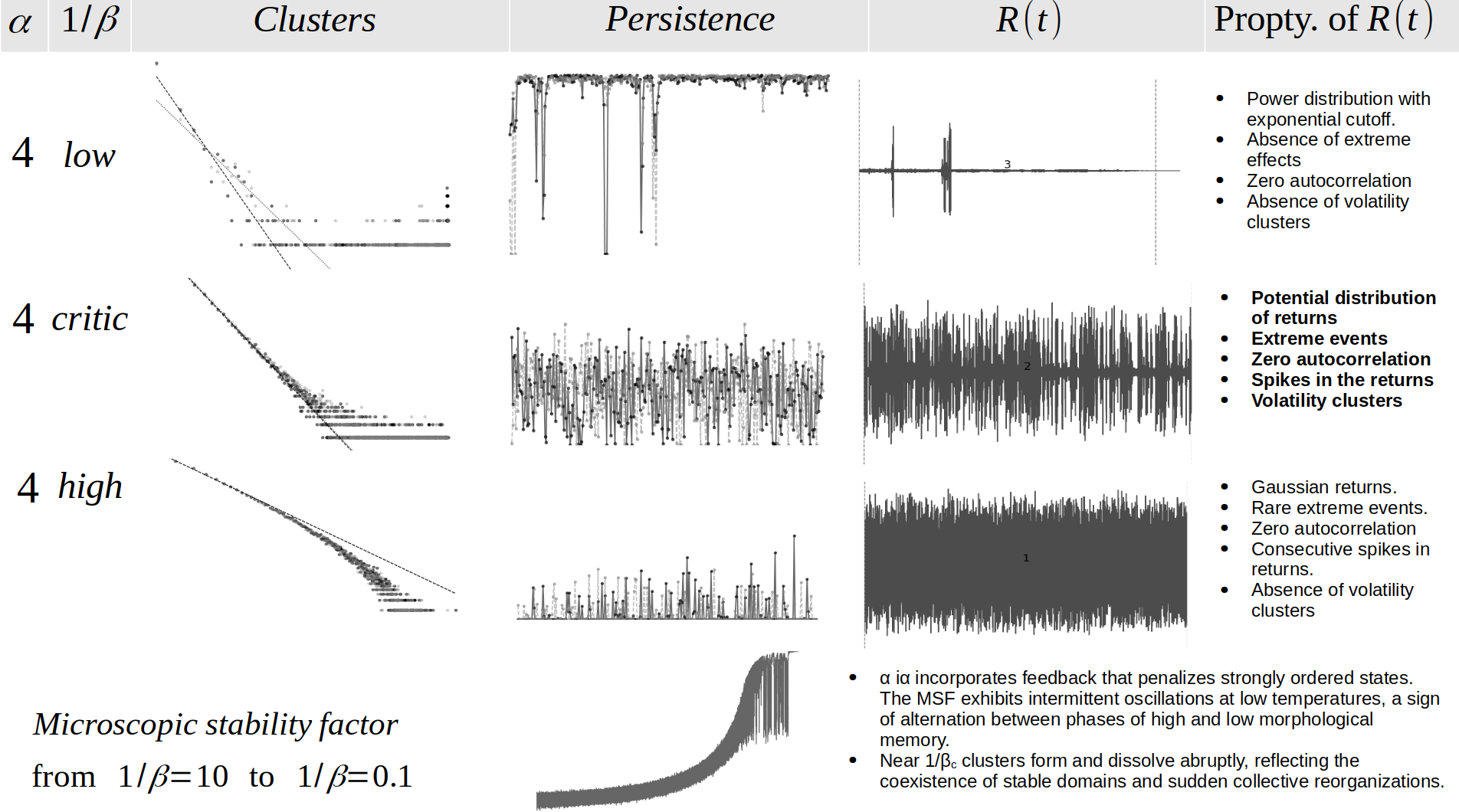} 
\par\end{centering}
\caption{Summary of the elements that determine the distribution of returns
$R(t)$, for different temperatures with $\alpha=4$.}
\end{figure}

\section{Apendix A: $\alpha=0$ }

\subsection{Low $1/\beta$ regimen:}

~

In this regime ($1/\beta\ll$ $1/\beta_{c}$), the system naturally
tends toward ferromagnetic order due to the dominance of the neighbor
interaction J. However, by setting $\alpha=0$, the feedback term
proportional to $\left|M(t)\right|$ disappears, eliminating the penalty
on the global magnetization. This results in a faster evolution toward
full alignment: in the first steps (0), the spins are disordered,
and immediately in steps 50, 100, and 290 they align completely, defining
larger domains. The absence of the global coupling term means that
the magnetization acts more rapidly and achieves ferromagnetic stabilization
almost immediately.

\begin{figure}[H]
\includegraphics[width=\linewidth]{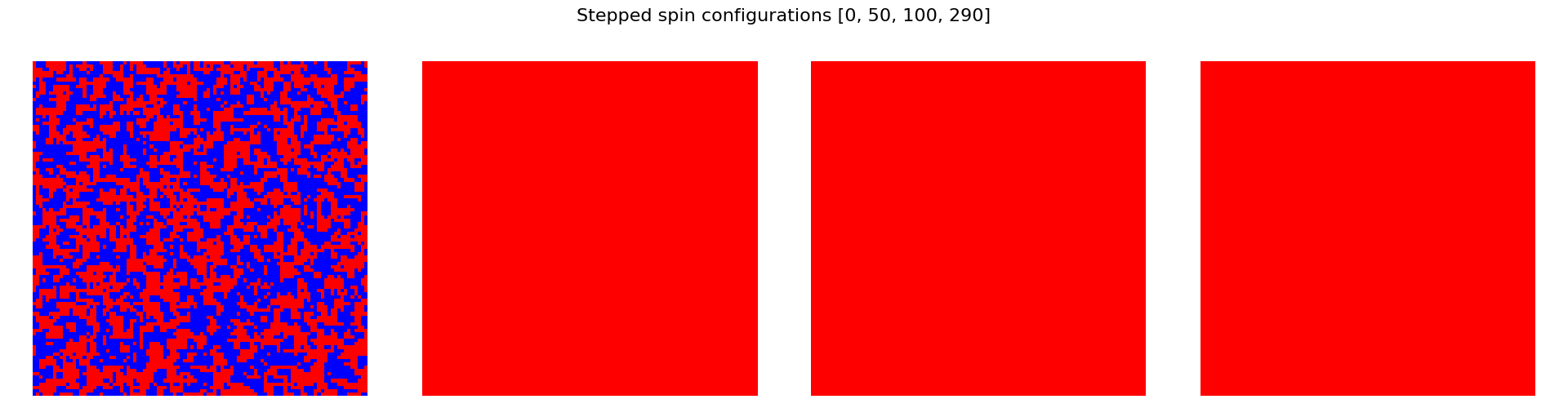}

\caption{Spin configurations in the low-temperature regime for $\alpha=0$.
The formation of coherent domains is observed as the dynamics progress
(Color online).}
\end{figure}

\subsection{Regimen $1/\beta_{c}$:}

~

As the value of $1/\beta_{c}$ approaches, the configurations begin
in a mostly disordered state, without a well-defined cluster structure.
Near $1/\beta_{c}$, domains of varying sizes emerge, with small disordered
regions coexisting with larger areas of aligned spins that begin to
dominate the lattice. For $\alpha=0$, the dynamics depend solely
on the local coupling between the four neighbors, without global feedback
to stabilize the magnetization. In the intermediate panels (50, 100,
and 290), the competition between the two types of domains---one
dominant and the other lagging---is evident, reflecting the coexistence
of phases and the proximity to the critical behavior of the Ising
model without an effective global field.

\begin{figure}[H]
\begin{raggedright}
\includegraphics[width=\linewidth]{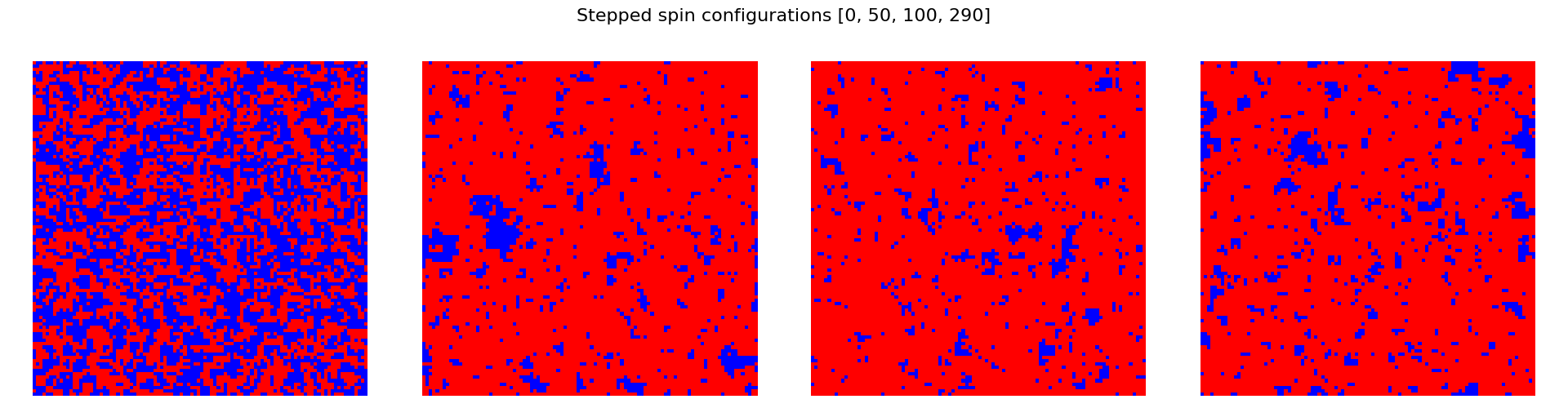} 
\par\end{raggedright}
\caption{Spin configurations in the vicinity of the critical temperature for
$\alpha=0$. The coexistence of multi-scale domains is observed, characteristic
of the critical point of the Ising model without a global field (Color
online).}
\end{figure}

\subsection{High $1/\beta$ regimen:}

~

In the high-temperature regime ($1/\beta\gg$ $1/\beta_{c}$), thermal
agitation completely dominates the dynamics. The configurations at
different steps exhibit a persistent random pattern, with no defined
domains or appreciable spatial correlation. The absence of a global
term ($\alpha=0$) eliminates any tendency toward collective alignment,
and the thermal fluctuations far exceed the coupling energy between
neighbors. Consequently, the average magnetization is zero, and the
system behaves like a disordered spin gas, where small clusters are
ephemeral and quickly disappear. 
\begin{flushleft}
\begin{figure}[H]
\begin{raggedright}
\includegraphics[width=\linewidth]{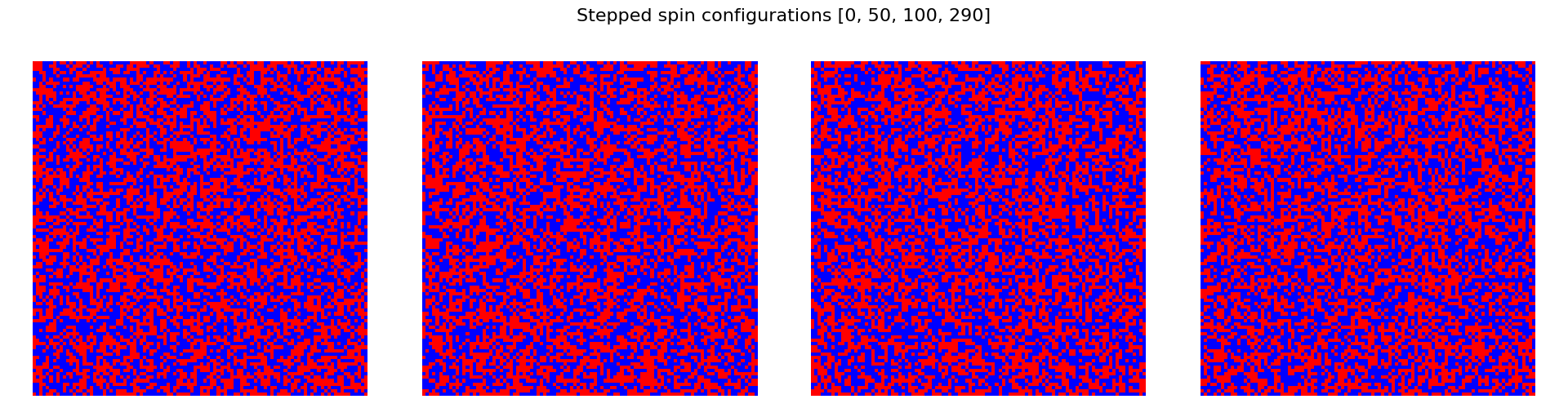} 
\par\end{raggedright}
\caption{Spin configurations in high-temperature regime for $\alpha=0$. The
system remains disordered due to thermal dominance and the absence
of global coupling, evidencing the typical paramagnetic state of the
Ising model (Color online).}
\end{figure}
\par\end{flushleft}

~

\end{document}